The Peculiar Photometric Properties of 2010 WG9: A Slowly-Rotating Trans-Neptunian Object from the Oort Cloud

David Rabinowitz[1], Megan E. Schwamb[1], Elena Hadjiyska[1], Suzanne Tourtellotte[2], Patricio Rojo[3]
[1]Yale University, Center for Astronomy and Astrophysics, New Haven, CT, USA
[2]Yale University, Astronomy Department, New Haven, CT USA
[3]University of Chile, Dept. of Astronomy, Santiago, Chile

ABSTRACT

We present long-term BVRI observations of 2010 WG9, an ~100-km diameter trans-Neptunian object (TNO) with an extremely high inclination of 70 deg discovered by the La Silla - QUEST southern sky survey. Most of the observations were obtained with ANDICAM on the SMARTS 1.3m at Cerro Tololo, Chile from Dec 2010 to Nov 2012. Additional observations were made with EFOSC2 on the 3.5-m NTT telescope of the European Southern Observatory at La Silla, Chile in Feb 2011. The observations reveal a sinusoidal light curve with amplitude 0.14 mag and period 5.4955 +/- 0.0025d, which is likely half the true rotation period. Such long rotation periods have previously been observed only for tidally-evolved binary TNOs, suggesting that 2010 WG9 may be such a system. We predict a nominal separation of at least 790 km, resolvable with HST and ground-based systems. We measure B-R = 1.318 +/- 0.029 and V-R = 0.520 +/- 0.018, consistent with the colors of modestly red Centaurs and Damocloids.  At I-band wavelengths, we observe an unusually large variation of color with rotational phase, with R-I ranging from 0.394 +/- 0.025 to 0.571 +/- 0.044. We also measure an absolute R-band absolute magnitude of 7.93 +/- 0.05 and solar phase coefficient 0.049 +/- 0.019 mag/deg.

1. INTRODUCTION

   Here we present photometric observations of 2010 WG9, a trans-Neptunian object (TNO) with an unusually high inclination discovered by the La Silla – QUEST southern-sky survey for TNOs  (Rabinowitz et al. 2012). With perihelion q = 18.8 AU, aphelion Q = 88.3 AU, and inclination i = 70.2 deg, 2010 WG9 is one of only five known solar system bodies with  i > 60 deg and q > 15 AU, including the retrograde object 2008 KV42 (Gladman et al. 2009). Brasser et al (2012a) refer to these bodies as high-inclination, high-perihelion (HiHq) objects. Because they have dynamical lifetimes of ~200 Myr limited by encounters with Neptune and Uranus, there must be a reservoir resupplying them. Brasser et al. show that the Oort cloud is the most likely source. On Gyr timescales galactic tides can lower the perihelia of some Oort cloud bodies to the point where they are captured by Uranus and/or Neptune and thereby removed from the Oort cloud. Thus, some HiHq objects may derive from the same inner Oort-cloud population revealed by Sedna and 2000 CR105 (Fernandez and Brunini 2000, Brasser et al. 2012b). The likelihood for the migration is extremely small (~$10^{-5}$), explaining why HiHq bodies were not predicted in earlier numerical simulations (e.g. Levison, Dones, & Duncan 2001). Nonetheless there are enough Oort cloud bodies to supply the observed population. The alternative possibility, that HiHq bodies migrate from the Kuiper Belt through planetary interactions alone, is vanishingly small.

   Observations of 2010 WG9 thus present a unique opportunity to measure the physical properties of a likely returning member of the Oort-cloud before it has reached the inner solar system.  Unlike long-period and Halley-family comets, Damocloids, and other high-inclination objects with perihelia interior to Saturn's orbit, 2010

WG9 has likely never approached the Sun closer than Uranus since its ejection into the Oort cloud early in the solar system's history. Furthermore, the body has likely spent most of the age of the solar system at distances larger than ~1000 AU, where volatile loss rates due to solar heating are negligible. In the Kuiper Belt, on the other hand, it is expected that all but the largest bodies have lost their surface volatiles owing to solar heating (Schaller & Brown 2007). It is therefore possible that 2010 WG9 retains a pristine surface composition compared to most TNOs and to previously observed bodies of likely Oort-cloud origin.

Here we present measurements of the brightness of 2010 WG9 in B, V, R, and I measured at several different epochs, and long-term I- and R-band observations revealing the rotation period, color, and slope of the solar phase curve. The resulting light curve reveals several physical properties rarely observed for outer solar system bodies, including an extremely slow rotation and large variations in color across the surface. To put the observed color in context, we compare the mean colors we observe for 2010 WG9 to other bodies on HiHq orbits and to other bodies that have likely migrated from the Oort cloud into the inner Solar System. We also compare to the Centaurs, which are generally representative of the Kuiper Belt. These comparisons probe the diversity of colors for Oort-cloud bodies, previously constrained only by observations of the devolatilized cores of long-period comets and of the few known Sedna-like bodies. Our color comparisons also test for any differences between the HiHq bodies and the Kuiper Belt objects (KBOs) which might result from differences in their regions of formation in the inner solar system. We also discuss the implications of the unusual rotation and color variability assuming 2010 WG9 is a tidally-evolved binary.

2. OBSERVATIONS

2.1 instrumentation

Most of the observations we report here were obtained in service mode by on-site operators at Cerro Tololo, Chile using the optical channel of the A Novel Dual Imaging Camera (ANDICAM) on the Small and Moderate Aperture Research Telescope System (SMARTS) 1.3 m telescope. The optical imager is a Fairchild 2K x 2K CCD with a pixel scale of 0.37". Table 1 summarizes the observing circumstances in each of three observing runs. All of the SMARTS observations consist of single 10-minute exposures in Johnson/Cousins B, V, R, and I. Observations made 2010 Dec 3 and 4 were taken in photometric conditions. Additional pairs of R-band images were taken nearly every other night from 2010 Dec 22 to 2011 Mar 26 in mixed conditions. Finally, pairs of both R- and I-band data were taken every night from 2012 Nov 10 to 20, also in mixed conditions. Seeing typically ranged from 1.0" to 2.0". All images were pre-processed using standard bias and twilight flat fields.

For more precise photometry, we made observations in classical mode with the 3.5-m New Technology Telescope (NTT) at the European Southern Observatory (ESO) at La Silla, Chile using the ESO Faint Object Spectrograph and Camera (EFOSC2). The imaging array consists of a thinned, 2Kx2K Loral/Lesser CCD with 15-um pixels. Exposures were taken 2x2 binned, yielding a pixel scale of 0.24". We recorded exposures in Bessel B, V, R, and Gunn i in photometric conditions on 2011 Feb 5 and 6, with seeing of 0.6" – 0.8" on Feb 5 and 1.1" - 1.3" on Feb 6. We recorded additional R- and B-band observations spanning 2.6 h on 2011 Feb 8 in non-photometric conditions (thin cirrus) with seeing ranging from 0.9" to 1.3". Target exposure times on all three nights ranged from 60 to 600 seconds. All observations were pre-processed using bias fields and twilight flats. Note that on Feb 8, R-band observations of the same target field were made in service mode with the SMARTS 1.3m at Cerro Tololo, where the conditions were photometric.

2.2 SMARTS calibrations

We calibrated all of the SMARTS data we obtained prior to 2012 using our standard reduction procedure described in detail by Rabinowitz et al. (2006, 2007). In this method we first choose bright field stars present in all the target images observed on photometric nights and calibrate them with respect to Landolt standards. We then calibrate the target in each exposure (including observations on non-photometric nights) by measuring the target flux relative to the calibrated field stars in the same image. To optimize signal to noise, we use a small aperture for these relative flux measurements (diameter 2.2", slightly larger than the typical seeing). We use the APHOT routines in IRAF to make these measurements. Our typical systematic error using this method is ~1.5%.

We used a new method to calibrate the 2012 SMARTS data. Instead of calibrating secondary standards in each field, which would have required additional observations on photometric nights, we used the UCAC4 catalog (Zacharias et al. 2013) to identify stars with known magnitudes in Johnson B and V and Gunn g, r, and i in each target field. Note that the UCAC4 catalog has incorporated the results of the recently-completed AAVSO Photometric All-Sky Survey (Henden et al. 2012), which has covered nearly the whole sky in BVgri down to mag V~16. We found that all of our 2012 SMARTS exposures contain at least three UCAC4 stars with magnitudes faint enough to be unsaturated in our 10-min exposures, yet bright enough to have precisely measured magnitudes. For these calibrations we use SExtractor to determine target and field star magnitudes (using a small aperture of 2.2" diameter) rather than relying on IRAF routines. We also convert the cataloged Gunn g, r, and i values listed by UCAC4 to Johnson R and I using the following conversion formulas known to be accurate to ~1% for most stars (see http://www.sdss.org/dr5/algorithms/sdssUBVRITransform.html) :

$$R = r - 0.1837*(g-r) - 0.0971$$
$$I = r - 1.2444*(r-i) - 0.3820$$

For most of the 2012 SMARTS fields, we find that the above procedure yields zero-point calibrations with precision better than 5%.

2.3 NTT calibrations

To calibrate our photometric NTT observations (2011 Feb 5 and 6), we used standard regression methods to determine the difference between cataloged and instrumental magnitudes of observed Landolt standards in each band pass as a linear function of air mass and source color. From multiple observations of many different standards at widely varying air mass, we derived solutions yielding observed minus predicted magnitudes with variance of 1.4 to 2.7% on Feb 5 and 1.0 to 1.2 % on Feb 6. We then measured the flux of our target within small apertures (diameters 0.72" and 1.2" on Feb 5 and 6, respective), adjusted these fluxes using an aperture correction determined separately for each image from the field stars, and used our calibration to determine the resulting magnitudes in each filter.

To analyze our non-photometric NTT observations (2011 Feb 8), all of which were exposures of the same field over a time span of 0.11 days, we reprocessed the images to optimize the precision for relative photometry. The purpose was to remove the signal from very faint stars and galaxies near the moving location of the target. Although no such sources are obvious in the images, their presence at a faint level comparable to the noise in the sky background limits our photometric precision. The reprocessing procedure, performed separately for the B and R band images, was as follows.

1. We registered the images with respect to the field stars (linear pixel shifts) and scaled the images to have unity value for the mean sky background. This transformed our initial set of bias- and flat-field corrected images to a normalized set.

2. We median-averaged the normalized images to obtain a single, deep exposure of the field in each pass-band. Because the target moved 16.5" over the time span of the image set, the target signal was removed by the median averaging.

3. We divided each normalized image by the median image to obtain a set of sky-divided images from which the signal from the sky, field stars, and galaxies was largely removed. The only signal remaining in each image was that of the target.

4. For each target observation, we used a fixed small aperture (1.9" diameter, comparable to the seeing) to measure the target magnitude in the sky-divided image, and also to measure the average magnitude of a fixed set of field stars in the normalized image. Any variations in the target signal resulting from changes in the PSF or sky transparency would affect the brightness of the target and field stars equally.

5. We subtracted the average field-star magnitude from the target magnitudes to obtain a relatively calibrated light curve

After using the above procedure to obtain a relative calibration, we were able to absolutely calibrate the R-band observations with respect to secondary standards in the field. We were able to calibrate these field standards using our SMARTS observations of the same field that had been incidentally taken the same night in photometric conditions. Unfortunately, we could not calibrate the B-band magnitudes of the field stars because we did not obtain a photometric B-band observation of the field. Hence the resulting B-band magnitudes, while precisely calibrated relative to one another, can not be referred to an absolute scale. To make the observations useful for measuring rotational variability in the R-band, we have added a zero point so that their mean value matches the mean for the R-band observations observed the same night (see discussion of Fig. 4 in Sec. 3.1).

Table 2 lists the resulting brightness measurements (M') and their measurement error ($\sigma$) from all of our SMARTS and NTT observations, excluding target measurements contaminated by cosmic ray hits or nearby bright stars. We also exclude R-band observations with $\sigma > 0.1$ mag. Table 2 also lists the Julian Date (JD) at the mid-time of each exposure (t), the solar phase angle ($\alpha$), the target distance to the Sun ($r_s$) and to the Earth ($r_e$), the reduced magnitude (M = M' – 5log[$r_s r_e$] ), and the light travel time ($\Delta t$) relative to the first observation in the table. In the proceeding analysis of the light curve, all observations times have been adjusted by $\Delta t$ to account for the relative motion of the target and Earth over the long time span of the observations. The largest time correction is ~10 minutes. The B-band NTT observations from 2011 Feb 8, adjusted to align with the R-band light curve, appear in Table 2 with label "$R_B$".

3. RESULTS AND ANALYSIS

3.1 The rotational light curve

Figures 1a - 1d show our measured values of the reduced magnitude in the B, V, R, and I bands versus JD. In the absence of rotational modulation and solar phase dependence, each of these light curves would be time dependent. The plot is separated into 4 panels to highlight the observations corresponding to the different observing campaigns listed in Table 2. Note that the time span is very different for each panel. Panel a shows only the SMARTS results obtained 2010 and 2011 (time span 100 days). Panel b shows all of the NTT results (time span 4 days) and the single R-band observation obtained concurrently using SMARTS on 2011 Feb 8. Panel c shows an expanded view of the NTT observations obtained 2011 Feb 8 alone (the relatively calibrated B-band observations, shifted to overlay the R-band observations, are shown as small points in blue). Panel d shows the SMARTS observations obtained 2012 Nov 10-20. Note that we exclude the B-band observations obtained 2011 Feb 8 from the analysis presented below, except where specifically cited.

Figs 1a-d show that the observations have significant scatter with peak-to-peak amplitude ~0.3 mag. For each band-pass, Table 3 shows the number of observations, N, and the residual $\chi^2$ for the distribution (see column labeled "$\chi_o^2$"). Specifically, we evaluate the following expression,

$$\chi_o^2 = \Sigma_i [M_i - <M>]^2/ \sigma_i^2 \tag{1}$$

The sum is over all observations, i = 1 to N, with magnitude, $M_i$, measurement error, $\sigma_i$, and where <M> is the weighted average magnitude,

$$<M> = \Sigma_i [M_i /\sigma_i^2]/\Sigma_i [1/\sigma_i^2] \qquad (2)$$

The resulting $\chi^2$ values are much larger than the number of degrees of freedom, N-1. This is particularly evident for the R-band observations, where $\chi^2$ = 232 with N = 49. This indicates significantly more variation than would be expected by chance for a source with no intrinsic variability. The scatter appears only when comparing observations taken days apart (Fig. 1a, 1b, and 1d), but not over an interval of a few hours (Fig 1c). The likely cause of the variation is rotational modulation with a period of at least a few days.

There is also a long-term $\alpha$-dependence to the observations. Fitting a linear relation between R-band magnitude and $\alpha$ (which varies from 1.7 to 2.9 deg), we measure a slope $\beta$ = 0.053 +/- 0.020 mag/deg. This variation is significant, but cannot explain the day-to-day variations we observe. Subtracting the solar phase dependence reduces the R-band $\chi^2$ from 232 to 212, but this is still significantly larger than the number of degrees of freedom (N-2 = 47). In Section 3.2 we re-derive $\beta$ after fitting and subtracting rotational variability. This changes the result slightly, but not significantly. For the purposes of measuring variability, we adopt the above value as it simplifies the analysis.

To search for a periodicity, we first removed the measured phase-angle dependence from the R-band observations plotted in Fig 1a-1d. To the resulting light curve we then we fit a sinusoidal function of time, F(t), with amplitude, A, and period, P, given by

$$F(t) = A\sin(2\pi [\phi + \omega(P,t)]) \qquad (3)$$

where $\omega(P,t)$ is the rotational phase, $\omega(P,t) = (t-t_o)/P$, at time, t, with respect to chosen epoch, $t_o$, and $\phi$ is a rotational phase offset. To obtain the best fit values for parameters P, A, and $\phi$, we then iteratively minimized $\chi^2$ given by

$$\chi^2 = \Sigma_i [M_i – <M> - F(t_i)]^2/ \sigma_i^2 \qquad (4)$$

The minimum value was found by varying A from 0.1 to 0.2 mag by increments of 0.01 mag, varying $\phi$ over its full range, 0 to 1, by increments of 0.01, and varying P from 0.05 to 20 d in uniform logarithmic increments, $\log[20.0/0.05]/10^5$. This search range for A comfortably brackets the amplitudes we would expect given the scatter in the observations. The search range for P covers the minimum period expected for a strengthless rubble pile (~0.08 d) up to the maximum capable of producing the day-to-day variations we observe (~10 days). Epoch $t_o$ was fixed at JD=2455500.

Figure 2 shows the resulting plot of $\chi^2$ vs P, where $\chi^2$ is the minimum value found for all values of A and $\phi$. It is apparent that there is one conspicuous minimum, $P_L$ = 5.4955 d, where $\chi^2$ = 64.4. Given the large reduction in $\chi^2$ compared to $\chi_o^2$, it is clear that there are significant correlations in the R-band light curve with this period. We obtain the best fit with A = 0.144 mag and $\phi$ = 1.27. Fixing A at this best value, we then estimate an uncertainty of +/- 0.0025 d for P by finding the range of periods centered on $P_L$ such that the minimum $\chi^2$ (for all values of $\phi$) exceeds $\chi^2$ = 64.4 by less than 2.3. This yields the 1-sigma uncertainty when two degrees of freedom ($\phi$ and P) are allowed to vary.

There is another $\chi^2$ minimum at $P_S$ = 1.2226 +/- 0.0001 d, where $\chi^2$ = 91.8 (with A = 0.130 mag and $\phi$ = 4.30). Given the much larger $\chi^2$ for this secondary minimum compared to $P_L$, it is clear that a sinusoid with this period does not fit the data as well. Also, one of these periods is clearly an alias of the other because $P_L/P_S$ = 4.504 is very close to a small integer ratio, 9/2. The ambiguity results from the one- and two-day sampling intervals of the SMARTS observations in 2010 and 2011, which make up the bulk of the R-band observations. It is likely that $P_L$ is the correct period and $P_S$ is the alias. However, in both cases the $\chi^2$ value is significantly larger than would be expected by chance given the number of degrees of freedom (46) in the sinusoidal fit (the likelihood is less than 5% for $\chi^2 > 62$). This implies that the measurement errors are larger than we have estimated, or that a simple sinusoid is not the best fit to the light curve. For example, adding a 3% systematic error in quadrature to each of our SMARTS R-band observations lowers the best-fit $\chi^2$ to 56, which has a chance likelihood of 15%. Below we give consideration to both period solutions and provide a detailed comparison of the model light curves to the observations in order to decide the correct value.

Figures 3a and 3b show the phased rotational light curves we obtain in B, V, R, and I when we assume $P = P_L$ and $P_S$, respectively. Each curve in each figure is a plot of the same data we show in Fig. 1, but with JD replaced by the fractional part of the rotational phase, $\omega_i$. The light curves are repeated over two periods to show continuity. Also, all the observations have been adjusted to remove the measured R-band $\alpha$ dependence, as discussed above. For each period, the best-fit R-band sinusoid, $F(t)$, is plotted as a solid line. For comparison with each of the other band passes, $F(t)$ is re-plotted with an offset matching the measured values for B-R, V-R, and I-R (see below). Table 3 lists the $\chi^2$ values we obtain for each light curve after subtracting $F(t)$ (see columns labeled "$\chi_L^2$" and "$\chi_S^2$"). These were computed using Eqs. (1) and (2), but with $M_i$ replaced by $M_i-F(t_i)$.

Comparing Figs. 3a and 3b, it is not obvious by visual inspection which period yields the best fit. In both cases, $F(t)$ faithfully follows the phased R-band oscillations, and the B and V light curves are also well fit. This fitting of the B and V observations by an R-band light curve shows that the observed oscillations are true rotational modulations rather than measurement artifacts. The $\chi^2$ values listed in Table 3 for the B and V light curves are significantly reduced after subtracting $F(t)$ for both period solutions. The new values are now comparable to the number of degrees of freedom (N-1 = 3, for both bands), again validating the R-band fits. For neither period does $F(t)$ correlate well with the I-band light curve. After subtracting $F(t)$, the I-band variance increases, resulting in a larger $\chi^2$. Thus it appears that the I-band rotational light curve differs from the light curve at bluer wavelengths, an indication of possible variations of color with rotational phase.

Figures 4a and 4b offer a more decisive comparison of the two period solutions. These figures again show the phased R-band data (red points) from Figs 3a and 3b, respectively, but now restricted to the NTT measurements obtained 2011 Feb 8 which span 2.6 hours and achieve the highest precision. The best-fit sinusoids from Figs. 3a and 3b are again plotted as solid lines. We also show the B-band observations (small blue points), adjusted so that their mean matches the R-band mean over this restricted phase range. Although this sequence covers only a small range in rotation phase, it was fortuitously recorded at the mid-range of the light curve where rotational brightness variations are most rapid. Note that for $P_L$ we expect a brightness slowly increasing with phase, whereas for $P_S$ we expect an intensity diminishing with phase at a relatively rapid rate. The residuals after subtracting the fits are plotted in Figs. 4c and 4d, respectively.

Comparing Figs 4a and 4b, it is now clear that $P_L$ provides a much better fit than $P_S$. Whereas the R-band observations alone or in combination with the B-band are consistent with the slow rise in brightness expected from $P_L$, they do not support the rapidly diminishing brightness expected for $P_S$. The $\chi^2$ after subtracting $F(t)$ is 12.5 for $P_L$, with N-1= 9 degrees of freedom. This scatter is only slightly larger than expected given the measurement errors, with a formal likelihood of ~20% that observations are consistent with $F(t)$. For $P_S$, on the other hand, we have $\chi^2 = 40$ which has vanishing likelihood (< 0.01%). Hence we consider $P_L$ to be the only valid period fitting the observations.

Our observations are not precise enough to determine if the number of peaks in the light curve is one per rotation, as expected from a single albedo spot on a spherical body, or two, as expected from the variation in the projected area of an elongated body. Bodies of ~100 km diameter such as 2010 WG9 generally have rotational light curves with higher-amplitude oscillations than larger bodies (Sheppard, Lacerda, & Ortiz 2008, Benecchi & Sheppard 2013). This is indicative of greater asphericity for the smaller bodies, possibly due to past collisional impacts. It is therefore likely that the variability we see for 2010 WG9 is mostly from changes in the projected area and the true rotation period is twice what we have measured, or 10.991 d.

3.2 Colors and solar phase coefficient

Having found a good fit to the rotational light curve, we can now derive a rotation-corrected value for $\beta$ and for the absolute magnitude, $H_R$, in the R band. We do this by linear regression, adjusting H and $\beta$ to minimize $\chi^2$ given by

$$\chi^2 = \Sigma_i \, [M_i - F(t_i) - (H + \beta \, \alpha_i)]^2 / \sigma_i^2 \qquad (5)$$

where $\alpha_i$ is the solar phase angle at time $t_i$ and $M_i$ is the reduced R-band magnitude. This yields $H_R$ = 7.93 +/- 0.05 mag and $\beta$ = 0.049 +/- 0.019 mag/deg, with $\chi^2$ = 56.9 and N-2 = 47 degrees of freedom. This new value for $\beta$ does not differ significantly from the value we derive earlier without rotation correction. For deriving colors, however, we adopt this new value as it has a marginal influence on the results.

To accurately measure the colors for 2010 WG9, we must account for both the rotational variability of the light curve and the $\alpha$ dependence. Two complimentary methods are available. The first is to separately determine the colors at each epoch where B, V, R, and I were all measured together over a short time interval compared to the variability. The results from these different epochs can then be averaged together to determine the mean colors. The second method is to correct the observations for their $\alpha$ dependence and then to subtract the R-band sinusoidal fit, F(t). With respect to the R band, we thus obtain the weighted average color, $\Delta M$, for each band pass using

$$\Delta M = \Sigma_i \, [M_i - F(t_i) - \beta \alpha_i]^2 / \sigma_i^2 \qquad (6)$$

In principle, this second method is a more accurate measure because $F(t) + \beta\alpha$ is a better estimate for the R-band magnitude at any given epoch than the individual R-band observations. We find that both methods give consistent results, and adopt the second because the results are more precise.

Table 4 lists the resulting colors, along with the number of measurements, N, used to determine the average and the residual $\chi^2$. Note that we list R-I for two ranges in rotational phase, R-$I_1$ for 0.0-0.4 and R-$I_2$ for 0.4-1.0. Close inspection of Fig 3a shows that the correlation between the I and R light curves is significant in both ranges, but there is a shift in R-I from one range to the other. The resulting $\chi^2$ values for B-R, V-R, R-$I_1$, and R-$I_2$ are all comparable to N-1, indicating good measures of the mean.

It is possible that the value we measure for $\beta$ may depend on wavelength. Using Eq. (5) above to separately measure $\beta$ for the B, V and I band passes, we derive $\beta$ = 0.082 +/- 0.083, 0.120 +/- 0.057, and -0.014 +/- 0.061 mag/deg, respectively. Given their large uncertainties, however, none of these $\beta$ values differs significantly from the R-band value. Many more observations in B, R, and I would be required to demonstrate a significant wavelength dependence. If there were a strong wavelength dependence, however, it would alter our color measurements by < ~0.1 mag. For example using $\beta$ = 0.082 mag/deg instead of $\beta$ = 0.049 mag/deg decreases B-R from 1.32 to 1.24 mag.

4. DISCUSSION

4.1 color

To put the colors we measure for 2010WG9 into context with similar solar system objects, we present in Fig. 5 a plot of R-I vs V-R for 2010 WG9 and for other inactive bodies that have likely returned from the Oort cloud. Medium-sized red diamonds represent two other HiHq TNOs (2008 KV42 and 2002XU9 from Sheppard 2010). Red triangles represent three distant bodies with exceptionally high aphelion or perihelion suspected to come from the inner Oort cloud (Sedna, 2006 SQ372, and 2000 OO67 from Sheppard 2010). Black squares represent the relatively nearby Damocloids thought to be devolatilized cores of Halley-type comets (from Jewitt 2005). Small green points represent the colors for Centaur asteroids listed in the MBOSS catalog (Hainaut et al. 2012). Unlike the other bodies represented in the figure, the Centaurs likely migrated from the Kuiper Belt without a passage through the Oort cloud. Their distribution of colors generally matches the colors of Kuiper Belt objects. Note that we represent 2010 WG9 in Fig. 5 by two different symbols (large unfilled and filled diamonds), corresponding to the two values we measure for R-I (rotational phases 0.0-0.4 and 0.4-1.0, respectively).

An examination of Fig. 5 shows that the colors we measure for 2010 WG9 do not distinguish it from Centaurs, Damocloids, or KBOs in general. The likely implication is that it has a similar surface composition, dominated by carbonaceous material with some fraction of polymerized organics. The presence or absence of surface water ice and/or volatile ices (e.g. methane, $CO_2$) cannot be ascertained from visual photometry alone. The object has the reddest spectral slope of the three known HiHq TNOs at V-band wavelengths (~0.6 um), and it has either the reddest or the bluest slope at R-band wavelengths (~0.7 um) depending on the rotational phase. It is interesting that the variation in R-I covers the whole range represented by the Damocloids. If the variance in red color among the Damocloids correlates with retention of a primordial organic, then 2010 WG9 may represent a body with a partially preserved and partially eroded surface. Other possible explanations for the variation in color are patchy areas of volatile ices (as on Pluto), or the presence of a large crater that has exposed relatively neutral-colored sub-surface material (water ice, or un-irradiated carbonaceous material).

Recently, Kiss et al (2013) report optical colors and thermal IR measurements of 2012 DR30, a borderline HiHq object with a = 1109 AU, i = 78 deg, q = 14.54 AU, and diameter 185 km. This body might have followed a followed a migration route from the Oort cloud like that of 2010 WG9, in which case it has since remained at distances beyond ~15 AU from the Sun. But it is equally likely that the object followed a migration route typical of long-period comets, with previous close encounters with Jupiter or Saturn pulling the orbit out of the Oort cloud. Hence, it may have experienced significant solar heating subsequent to its capture. The optical colors for 2012 DR30 are similar to 2010 WG9, except at blue wavelengths where the spectral slope is significantly flatter. On the basis of this feature and an absorption they observe in the z band, Kiss et al suggest that 2012 DR30 could be a V-type asteroid from the main belt. This interpretation could not apply to 2010 WG9, owing to its significantly larger perihelion and color differences.

4.2 rotation

The slow rotation period we measure for 2010 WG9 is extraordinary. Most known KBOs and asteroids have orbit periods of less than a day. Only a few dozen main-belt asteroids are known to have such very slow rotations (Masiero et al. 2009). Among these much smaller bodies, the slow rotation might be expected from the Yarkovsky effect (Pravec et al 2005). But such slowing is not expected for ~100-km diameter bodies at large distances from the sun, such as 2010 WG9. Of the ~100 TNOs and Centaurs with measured periods, the mean period is ~ 7h and all but a few have periods < 26 hours (Duffard et al 2009, Benecchi & Sheppard 2013). The only known distant bodies with rotation periods much longer than one day are the synchronously locked binary objects Pluto/Charon and Sila/Nunam (Grundy et al. 2012). This suggests that 2010 WG9 may be a similar tidally-evolved binary system.

Assuming 2010 WG9 is a system composed of two equal mass bodies on circular orbits about their center of mass, we can estimate their separation, Δ, based on an estimate of their combined mass, m. We start with the absolute V-band magnitude which we measure to be H = $H_R$ + V-R = 7.9 + 0.52 = 8.4. Known TNOs with comparable absolute magnitudes have measured diameters of ~100 km (Stansberry et al 2008). If the binary consists of two bodies of equal albedo and size, then their individual diameters are $100/2^{1/2}$ ~ 70 km. Scaling from Pluto's known mass (1.33 x $10^{22}$ kg) and diameter (2.4 x$10^3$ km) and assuming a density of 1 gm cm$^{-3}$ for the binary components (half the density of Pluto), we expect m ~ 3.3 x $10^{17}$ kg. Then using Kepler's law, we obtain Δ = 500 km for an orbital period equal to the measured period, $P_L$ = 5.4955 d, or Δ = 790 km if the orbit period is twice the measured period, $2P_L$ = 10.991 d. A known binary Centaur with comparable physical properties is the Typhon/Echidna system for which the component diameters are 112 and 56 km, the separation is ~1300 km (Stansberry et al. 2012), and the orbital period is 18.97 days (Grundy et al. 2008). The binary Trojan Patroclus-Menoetius is also similar, composed of nearly equal sized bodies of ~100 km diameter separated by 654 km with a synchronously locked rotation period of 4.29 d (Mueller et al 2010). At the current distance of 18.8 AU for 2010 WG9, a separation of 790 km corresponds to angular separation ~0.06". Such separations are resolvable with the Hubble Space Telescope or large ground-based telescopes with adaptive optics (Grundy et al. 2011).

If 2010 WG9 is indeed a tidally locked binary system, then the color variation we see with rotational phase could relate to different surface properties for the facing and opposing hemispheres of the two bodies. For example, Stern (2009) proposes that impact ejecta from closely separated binaries would coat the facing hemispheres of tidally locked partners over the age of the solar system. This is especially relevant for bodies with small separation-to-diameter ratios. For 2010 WG9, this ratio might be ~15, one of the smallest for any known binary system. Alternatively, the color variation might relate to the formation mechanism of the binary. For example, the surface of Haumea is known to have a dark red spot (Lacerda 2009 an 2010, Fraser & Brown 2009). Though this spectroscopic feature has not been linked to any specific topological features on the surface of Haumea, its existence may relate to surface heterogeneity stemming from the collision that likely generated Haumea's satellite system (Leinhardt et al. 2010).

5. CONCLUSIONS

We have presented photometric observations of 2010 WG9 revealing a very slow rotation period and a large variation in color with rotational phase. While the nominal color we observe is not unusual for a distant body, the slow rotation is extreme, observed previously only for tidally locked binaries in the Kuiper Belt. The object also happens to follow a very high-inclination orbit, suggesting that it is a member of the Oort cloud captured by close approach to Uranus or Neptune. Could the unique photometric properties relate to its peculiar orbit? The variation in color may relate to partial retention of primordial surface volatiles or a primordial crust of irradiated, organic material. Clearly, additional observations are required to determine if 2010 WG9 is a binary system and to better measure the variation in color with orbital phase. The future discovery of additional HiHq TNOs will also better determine their range of photometric properties and the physical properties of their parent population.

This work was funded by NASA planetary astronomy award NNX10AB31G. M.E.S. is supported by an NSF Astronomy and Astrophysics Postdoctoral Fellowship (AST-1003258). We also thank the hard-working SMARTS queue managers and observers and the support staff at La Silla.

Table 1. Observing Circumstances

| Telescope | Instrument | Filters | Dates | No. Nights | Mode | Conditions |
|---|---|---|---|---|---|---|
| SMARTS 1.3m | ANDICAM | B, V, I, R | 2010 Dec 3, 4 | 2 | Service | photometric |
| SMARTS 1.3m | ANDICAM | R | 2010 Dec 22 - 2011 Mar 26 | 44 | Service | mixed |
| NTT 3.5m | EFOSC2 | B, V, R, i | 2011 Feb 5, 6 | 2 | Classical | photometric |
| NTT 3.5m | EFOSC2 | B,R | 2011 Feb 8 | 1 | Classical | cirrus |
| SMARTS 1.3m | ANDICAM | I, R | 2012 Nov 10 - 20 | 11 | Service | mixed |

**Table 2. Photometric Observations of 2010 WG9**

| JD-2450000 | Δt (min) | M' (mag) | σ (mag) | M (mag) | α (deg) | $r_s$ (AU) | $r_e$ (AU) | Filter | Telescope |
|---|---|---|---|---|---|---|---|---|---|
| 5534.65033 | 0.00 | 22.131 | 0.082 | 9.303 | 1.843 | 19.567 | 18.798 | B | SMARTS-1.3 |
| 5534.65793 | 0.00 | 21.356 | 0.062 | 8.528 | 1.843 | 19.567 | 18.798 | V | SMARTS-1.3 |
| 5534.66553 | 0.00 | 20.809 | 0.053 | 7.981 | 1.843 | 19.567 | 18.798 | R | SMARTS-1.3 |
| 5534.67316 | 0.00 | 20.256 | 0.068 | 7.428 | 1.843 | 19.567 | 18.798 | I | SMARTS-1.3 |
| 5535.60153 | 0.00 | 21.994 | 0.088 | 9.166 | 1.839 | 19.568 | 18.798 | B | SMARTS-1.3 |
| 5535.60912 | 0.00 | 21.128 | 0.055 | 8.300 | 1.839 | 19.568 | 18.798 | V | SMARTS-1.3 |
| 5535.61672 | 0.00 | 20.766 | 0.056 | 7.938 | 1.839 | 19.568 | 18.798 | R | SMARTS-1.3 |
| 5535.62435 | 0.00 | 20.368 | 0.082 | 7.540 | 1.838 | 19.568 | 18.798 | I | SMARTS-1.3 |
| 5553.66254 | 0.34 | 20.654 | 0.137 | 7.819 | 1.914 | 19.584 | 18.839 | R | SMARTS-1.3 |
| 5553.71019 | 0.34 | 20.931 | 0.187 | 8.096 | 1.915 | 19.584 | 18.840 | R | SMARTS-1.3 |
| 5558.67902 | 0.55 | 20.641 | 0.055 | 7.803 | 1.982 | 19.589 | 18.865 | R | SMARTS-1.3 |
| 5558.73460 | 0.56 | 20.760 | 0.072 | 7.922 | 1.983 | 19.589 | 18.865 | R | SMARTS-1.3 |
| 5561.63859 | 0.70 | 20.943 | 0.070 | 8.102 | 2.028 | 19.591 | 18.883 | R | SMARTS-1.3 |
| 5561.67750 | 0.71 | 21.000 | 0.089 | 8.159 | 2.029 | 19.591 | 18.883 | R | SMARTS-1.3 |
| 5563.65215 | 0.81 | 20.499 | 0.084 | 7.657 | 2.062 | 19.593 | 18.896 | R | SMARTS-1.3 |
| 5563.66033 | 0.81 | 20.656 | 0.079 | 7.814 | 2.062 | 19.593 | 18.896 | R | SMARTS-1.3 |
| 5563.69481 | 0.82 | 20.766 | 0.064 | 7.924 | 2.063 | 19.593 | 18.896 | R | SMARTS-1.3 |
| 5565.64137 | 0.93 | 20.819 | 0.058 | 7.975 | 2.097 | 19.595 | 18.910 | R | SMARTS-1.3 |
| 5565.69402 | 0.93 | 21.044 | 0.094 | 8.200 | 2.098 | 19.595 | 18.911 | R | SMARTS-1.3 |
| 5567.64210 | 1.06 | 20.845 | 0.050 | 7.999 | 2.134 | 19.597 | 18.925 | R | SMARTS-1.3 |
| 5567.70360 | 1.06 | 20.889 | 0.057 | 8.043 | 2.135 | 19.597 | 18.926 | R | SMARTS-1.3 |
| 5570.59325 | 1.25 | 21.031 | 0.085 | 8.182 | 2.190 | 19.599 | 18.949 | R | SMARTS-1.3 |
| 5570.68273 | 1.26 | 20.930 | 0.079 | 8.081 | 2.191 | 19.599 | 18.950 | R | SMARTS-1.3 |
| 5573.58771 | 1.47 | 20.866 | 0.067 | 8.014 | 2.247 | 19.602 | 18.975 | R | SMARTS-1.3 |
| 5573.63733 | 1.47 | 20.874 | 0.059 | 8.022 | 2.248 | 19.602 | 18.975 | R | SMARTS-1.3 |
| 5577.67127 | 1.79 | 21.002 | 0.147 | 8.145 | 2.327 | 19.606 | 19.013 | R | SMARTS-1.3 |
| 5577.70409 | 1.79 | 20.932 | 0.135 | 8.075 | 2.327 | 19.606 | 19.014 | R | SMARTS-1.3 |
| 5581.60208 | 2.12 | 20.910 | 0.154 | 8.048 | 2.403 | 19.609 | 19.053 | R | SMARTS-1.3 |
| 5581.65115 | 2.13 | 20.831 | 0.161 | 7.969 | 2.404 | 19.610 | 19.054 | R | SMARTS-1.3 |
| 5589.60134 | 2.87 | 20.965 | 0.066 | 8.092 | 2.549 | 19.617 | 19.143 | R | SMARTS-1.3 |
| 5589.65287 | 2.87 | 20.869 | 0.067 | 7.996 | 2.550 | 19.617 | 19.144 | R | SMARTS-1.3 |
| 5591.61610 | 3.07 | 20.822 | 0.056 | 7.946 | 2.584 | 19.619 | 19.167 | R | SMARTS-1.3 |
| 5591.66986 | 3.07 | 20.700 | 0.058 | 7.824 | 2.585 | 19.619 | 19.168 | R | SMARTS-1.3 |
| 5597.56353 | 3.69 | 20.946 | 0.029 | 8.061 | 2.677 | 19.624 | 19.242 | R | NTT-3.5 |
| 5597.57321 | 3.69 | 21.436 | 0.027 | 8.551 | 2.677 | 19.624 | 19.242 | V | NTT-3.5 |
| 5597.57670 | 3.69 | 22.309 | 0.060 | 9.424 | 2.677 | 19.624 | 19.242 | B | NTT-3.5 |
| 5597.58002 | 3.69 | 20.417 | 0.047 | 7.532 | 2.677 | 19.624 | 19.242 | I | NTT-3.5 |
| 5597.58354 | 3.69 | 20.951 | 0.029 | 8.066 | 2.677 | 19.624 | 19.242 | R | NTT-3.5 |
| 5598.67190 | 3.81 | 21.087 | 0.030 | 8.200 | 2.693 | 19.625 | 19.256 | R | NTT-3.5 |
| 5598.67706 | 3.81 | 21.065 | 0.022 | 8.178 | 2.693 | 19.625 | 19.256 | R | NTT-3.5 |
| 5598.68125 | 3.81 | 21.570 | 0.029 | 8.683s | 2.693 | 19.625 | 19.256 | V | NTT-3.5 |
| 5598.68648 | 3.81 | 22.339 | 0.040 | 9.452 | 2.693 | 19.625 | 19.256 | B | NTT-3.5 |
| 5598.69171 | 3.81 | 21.059 | 0.026 | 8.172 | 2.693 | 19.625 | 19.256 | R | NTT-3.5 |
| 5600.53825 | 4.01 | 20.803 | 0.094 | 7.913 | 2.718 | 19.627 | 19.281 | R | SMARTS-1.3 |
| 5600.56864 | 4.02 | 20.854 | 0.015 | 7.964 | 2.719 | 19.627 | 19.281 | R | NTT-3.5 |
| 5600.57085 | 4.02 | 20.934 | 0.030 | 8.044 | 2.719 | 19.627 | 19.281 | $R_B$ | NTT-3.5 |
| 5600.57864 | 4.02 | 20.888 | 0.012 | 7.998 | 2.719 | 19.627 | 19.281 | R | NTT-3.5 |
| 5600.61215 | 4.02 | 20.896 | 0.012 | 8.006 | 2.719 | 19.627 | 19.282 | R | NTT-3.5 |
| 5600.61438 | 4.02 | 20.965 | 0.020 | 8.075 | 2.719 | 19.627 | 19.282 | $R_B$ | NTT-3.5 |
| 5600.62217 | 4.02 | 20.884 | 0.012 | 7.994 | 2.719 | 19.627 | 19.282 | R | NTT-3.5 |
| 5600.62439 | 4.02 | 20.958 | 0.023 | 8.068 | 2.719 | 19.627 | 19.282 | $R_B$ | NTT-3.5 |
| 5600.65358 | 4.03 | 20.908 | 0.012 | 8.018 | 2.720 | 19.627 | 19.282 | R | NTT-3.5 |
| 5600.65707 | 4.03 | 20.894 | 0.024 | 8.004 | 2.720 | 19.627 | 19.282 | $R_B$ | NTT-3.5 |
| 5600.66963 | 4.03 | 20.962 | 0.023 | 8.072 | 2.720 | 19.627 | 19.283 | $R_B$ | NTT-3.5 |
| 5600.67871 | 4.03 | 20.891 | 0.011 | 8.001 | 2.720 | 19.627 | 19.283 | R | NTT-3.5 |
| 5612.60374 | 5.40 | 20.711 | 0.147 | 7.801 | 2.843 | 19.638 | 19.448 | R | SMARTS-1.3 |
| 5612.64708 | 5.41 | 20.830 | 0.194 | 7.920 | 2.843 | 19.638 | 19.449 | R | SMARTS-1.3 |
| 5613.55909 | 5.52 | 20.841 | 0.091 | 7.929 | 2.850 | 19.639 | 19.462 | R | SMARTS-1.3 |
| 5613.65420 | 5.53 | 20.739 | 0.167 | 7.827 | 2.850 | 19.639 | 19.463 | R | SMARTS-1.3 |
| 5615.55986 | 5.76 | 21.058 | 0.081 | 8.143 | 2.862 | 19.641 | 19.491 | R | SMARTS-1.3 |
| 5615.62756 | 5.77 | 20.802 | 0.105 | 7.887 | 2.863 | 19.641 | 19.492 | R | SMARTS-1.3 |
| 5619.52076 | 6.24 | 20.825 | 0.099 | 7.903 | 2.881 | 19.645 | 19.548 | R | SMARTS-1.3 |
| 5631.57383 | 7.72 | 20.811 | 0.150 | 7.868 | 2.883 | 19.656 | 19.726 | R | SMARTS-1.3 |
| 5633.52340 | 7.96 | 21.180 | 0.128 | 8.234 | 2.876 | 19.658 | 19.755 | R | SMARTS-1.3 |
| 5633.58313 | 7.96 | 21.058 | 0.167 | 8.112 | 2.875 | 19.658 | 19.756 | R | SMARTS-1.3 |
| 5635.51741 | 8.20 | 20.644 | 0.146 | 7.695 | 2.866 | 19.660 | 19.784 | R | SMARTS-1.3 |
| 5635.58073 | 8.21 | 21.260 | 0.271 | 8.310 | 2.866 | 19.660 | 19.785 | R | SMARTS-1.3 |
| 5639.50821 | 8.68 | 20.818 | 0.205 | 7.862 | 2.841 | 19.663 | 19.842 | R | SMARTS-1.3 |
| 5639.57567 | 8.69 | 20.785 | 0.240 | 7.829 | 2.840 | 19.663 | 19.843 | R | SMARTS-1.3 |
| 5645.51194 | 9.40 | 20.991 | 0.118 | 8.025 | 2.787 | 19.669 | 19.928 | R | SMARTS-1.3 |
| 5645.55864 | 9.40 | 20.888 | 0.108 | 7.922 | 2.787 | 19.669 | 19.929 | R | SMARTS-1.3 |
| 5647.50265 | 9.63 | 20.924 | 0.101 | 7.955 | 2.766 | 19.671 | 19.956 | R | SMARTS-1.3 |
| 5647.54855 | 9.63 | 21.108 | 0.143 | 8.138 | 2.765 | 19.671 | 19.957 | R | SMARTS-1.3 |
| 6242.70228 | 6.55 | 21.078 | 0.099 | 8.079 | 1.931 | 20.315 | 19.587 | R | SMARTS-1.3 |
| 6243.64725 | 6.49 | 21.194 | 0.142 | 8.196 | 1.906 | 20.316 | 19.581 | R | SMARTS-1.3 |
| 6243.75179 | 6.49 | 21.024 | 0.086 | 8.026 | 1.903 | 20.316 | 19.580 | R | SMARTS-1.3 |
| 6244.67056 | 6.44 | 20.816 | 0.099 | 7.818 | 1.879 | 20.317 | 19.574 | R | SMARTS-1.3 |
| 6244.78920 | 6.43 | 20.948 | 0.070 | 7.950 | 1.876 | 20.318 | 19.573 | R | SMARTS-1.3 |
| 6244.83244 | 6.43 | 20.560 | 0.106 | 7.562 | 1.875 | 20.318 | 19.573 | I | SMARTS-1.3 |
| 6245.65362 | 6.38 | 20.528 | 0.176 | 7.531 | 1.854 | 20.319 | 19.568 | I | SMARTS-1.3 |
| 6245.70401 | 6.38 | 21.051 | 0.076 | 8.054 | 1.853 | 20.319 | 19.568 | R | SMARTS-1.3 |
| 6246.64226 | 6.33 | 20.595 | 0.142 | 7.598 | 1.828 | 20.320 | 19.562 | I | SMARTS-1.3 |
| 6246.68500 | 6.33 | 21.089 | 0.077 | 8.092 | 1.827 | 20.320 | 19.561 | R | SMARTS-1.3 |
| 6246.73468 | 6.33 | 20.737 | 0.120 | 7.740 | 1.826 | 20.320 | 19.561 | I | SMARTS-1.3 |
| 6246.79584 | 6.33 | 21.140 | 0.075 | 8.143 | 1.824 | 20.320 | 19.561 | R | SMARTS-1.3 |
| 6247.61837 | 6.28 | 21.146 | 0.086 | 8.150 | 1.803 | 20.321 | 19.556 | R | SMARTS-1.3 |
| 6247.67519 | 6.28 | 20.542 | 0.091 | 7.546 | 1.802 | 20.321 | 19.556 | I | SMARTS-1.3 |
| 6247.72594 | 6.28 | 21.209 | 0.070 | 8.213 | 1.800 | 20.321 | 19.555 | R | SMARTS-1.3 |
| 6247.78166 | 6.28 | 20.566 | 0.081 | 7.570 | 1.799 | 20.321 | 19.555 | I | SMARTS-1.3 |
| 6248.61765 | 6.24 | 20.603 | 0.173 | 7.607 | 1.778 | 20.322 | 19.550 | I | SMARTS-1.3 |
| 6248.66425 | 6.24 | 21.042 | 0.130 | 8.046 | 1.777 | 20.322 | 19.550 | R | SMARTS-1.3 |
| 6248.78505 | 6.23 | 21.193 | 0.075 | 8.197 | 1.774 | 20.322 | 19.549 | R | SMARTS-1.3 |
| 6249.76990 | 6.18 | 20.561 | 0.079 | 7.566 | 1.749 | 20.324 | 19.544 | I | SMARTS-1.3 |
| 6250.59534 | 6.15 | 20.753 | 0.100 | 7.758 | 1.728 | 20.325 | 19.540 | R | SMARTS-1.3 |
| 6250.64179 | 6.15 | 20.521 | 0.098 | 7.526 | 1.727 | 20.325 | 19.539 | I | SMARTS-1.3 |
| 6250.69644 | 6.14 | 20.821 | 0.056 | 7.826 | 1.726 | 20.325 | 19.539 | R | SMARTS-1.3 |
| 6250.75914 | 6.14 | 20.453 | 0.085 | 7.458 | 1.724 | 20.325 | 19.539 | I | SMARTS-1.3 |
| 6251.65839 | 6.10 | 21.015 | 0.089 | 8.021 | 1.702 | 20.326 | 19.534 | R | SMARTS-1.3 |
| 6251.69284 | 6.10 | 20.623 | 0.096 | 7.629 | 1.702 | 20.326 | 19.534 | I | SMARTS-1.3 |
| 6251.78558 | 6.10 | 20.632 | 0.096 | 7.638 | 1.699 | 20.326 | 19.534 | I | SMARTS-1.3 |
| 6252.59496 | 6.07 | 21.300 | 0.252 | 8.306 | 1.680 | 20.327 | 19.530 | R | SMARTS-1.3 |
| 6252.64225 | 6.07 | 20.895 | 0.186 | 7.901 | 1.679 | 20.327 | 19.530 | I | SMARTS-1.3 |
| 6252.70388 | 6.06 | 21.086 | 0.082 | 8.092 | 1.677 | 20.327 | 19.529 | R | SMARTS-1.3 |
| 6252.76114 | 6.06 | 20.629 | 0.106 | 7.635 | 1.676 | 20.327 | 19.529 | I | SMARTS-1.3 |

Table 3. $\chi^2$ for Measured Light Curves

| Bandpass | N | $\chi_o^2$ | $\chi_L^2$ | $\chi_S^2$ |
|---|---|---|---|---|
| B | 4 | 10.3 | 3.5 | 1.1 |
| V | 4 | 40.2 | 3.6 | 5.3 |
| R | 49 | 232.7 | 64.5 | 91.3 |
| I | 17 | 13.2 | 21.0 | 33.2 |
| $R_{NTT}$ | 10 | 12.5 | 14.7 | 34.9 |

Table 4. Average Colors for 2010 WG9

| Color | N | ΔM (mag) | Unc. (mag) | $\chi^2$ |
|---|---|---|---|---|
| B-R | 4 | 1.318 | 0.029 | 3.57 |
| V-R | 4 | 0.520 | 0.018 | 3.86 |
| R-$I_1$ | 4 | 0.571 | 0.044 | 0.51 |
| R-$I_2$ | 13 | 0.394 | 0.025 | 7.93 |

Notes: R-I is evaluated separately for rotational phase 0.0-0.4 and 0.4-1.0. See text for details.

Figure Captions

Figure 1. Reduced B, V, R, I magnitudes versus Julian Date for 2010 WG9 at four different epochs: (a) SMARTS observations from 2010 and 2011; (b) NTT observations from 2011 Feb 5, 6, and 8; (c) NTT observations from 2011 Feb 8; (d) SMARTS observations from 2012 Nov 10-20. Each band pass is represented by a different symbol: blue triangles (B), green hexagons (V), red squares (R), and black pentagons (I). In panel c, small blue points represent B-band data that has been relatively calibrated and then offset to best overlap the R-band data. These data are excluded from the analysis in the text, except where specifically cited.

Figure 2. A periodogram showing $\chi^2$ vs period for sinusoidal fits to the data. The best fit periods are found at $P_L$=5.4955 days and $P_S$=1.2226 days. A detailed analysis shows that $P_L$ is the correct period.

Figure 3. Rotationally phased light curves in B, V, R and I for two different periods: (a) $P_L$ = 5.4955 d and (b) $P_S$ = 1.2226 d. Each band pass is represented by a different symbol, as explained in Fig. 1. All light curves are repeated over two phase cycles to illustrate continuity across cycle boundaries. In each panel, the black curves are the best sinusoidal fit to the R-band data for the respective period. To match the B, V, and I band passes, R-band fit has been offset by the measured colors B-V, V-R, and I-R. Both periods appear to fit the B, V, and R measurements. Neither period provides a good fit to the I-band observations.

Figure 4. Rotationally phased light curves showing only the R and B band NTT observations from 2012 Feb 8. As in Fig 3, the black curves show the best sinusoidal fits. Panels (a) and (b) are for periods $P_L$ and $P_S$, respectively. Panels (c) and (d) show the residuals after subtracting the fits in panels (a) and (b), respectively. The spread in residual is smallest for period $P_L$.

Figure 5. R-I vs V-R for 2010 WG9 (large filled and unfilled red diamonds), other HiHq TNOs (medium-sized red diamonds), TNOs from the inner Oort cloud (red triangles), Damocloids (black squares), and Centaurs (small green points, with error bars not shown to avoid crowding of the figure). The colors for 2010 WG9 are consistent with all of these populations, except perhaps the few observed inner-Oort cloud bodies.

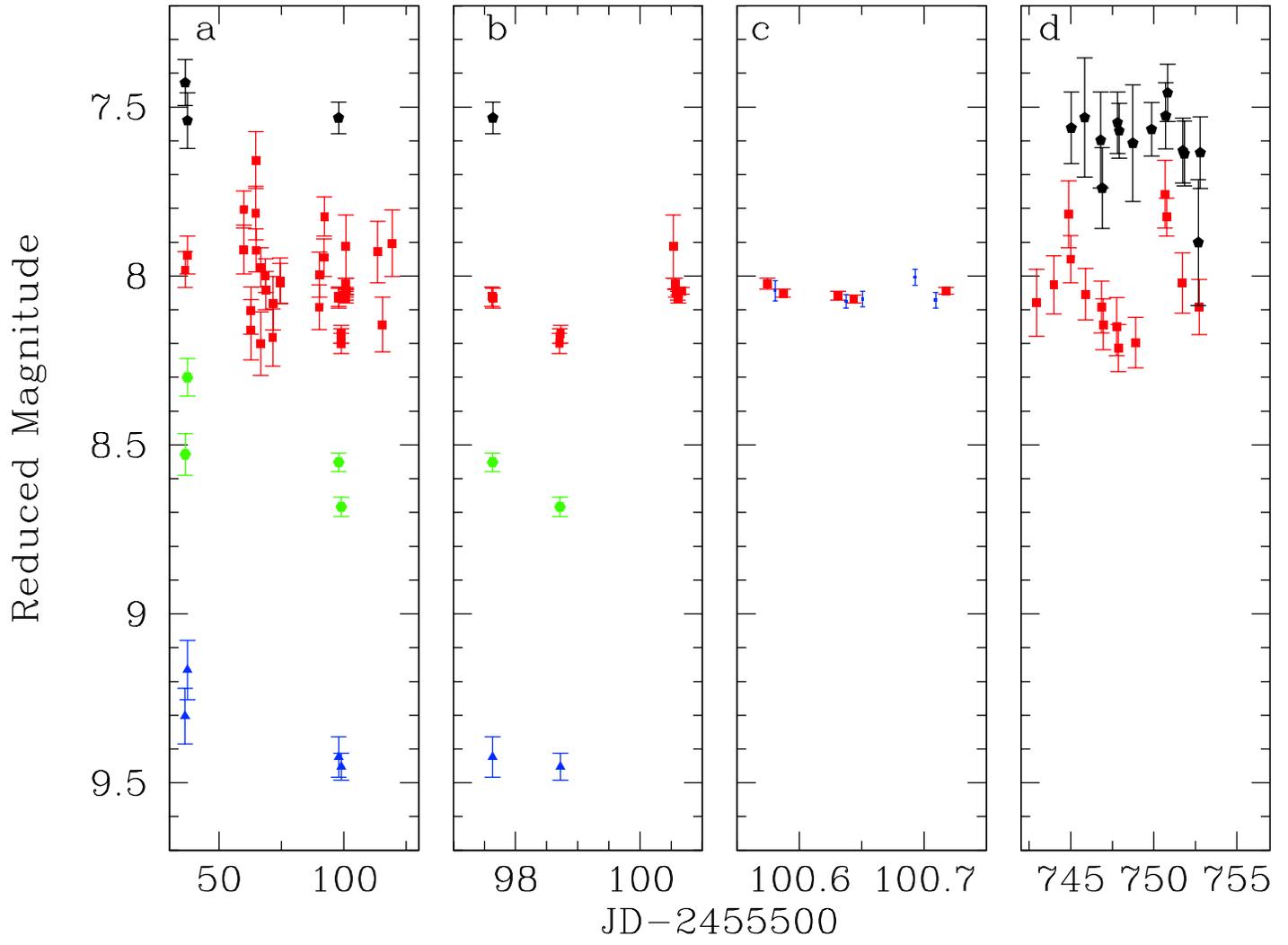

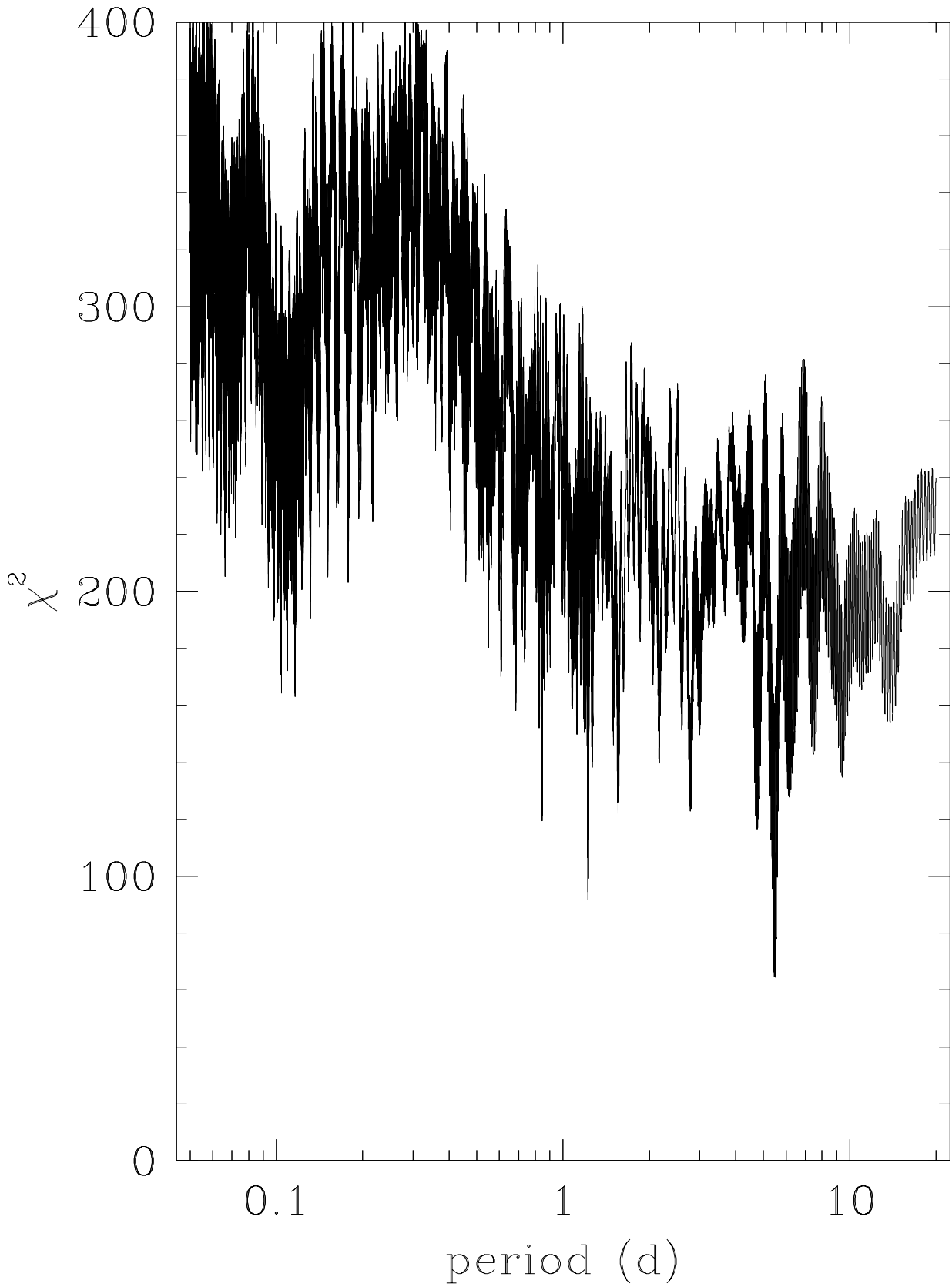

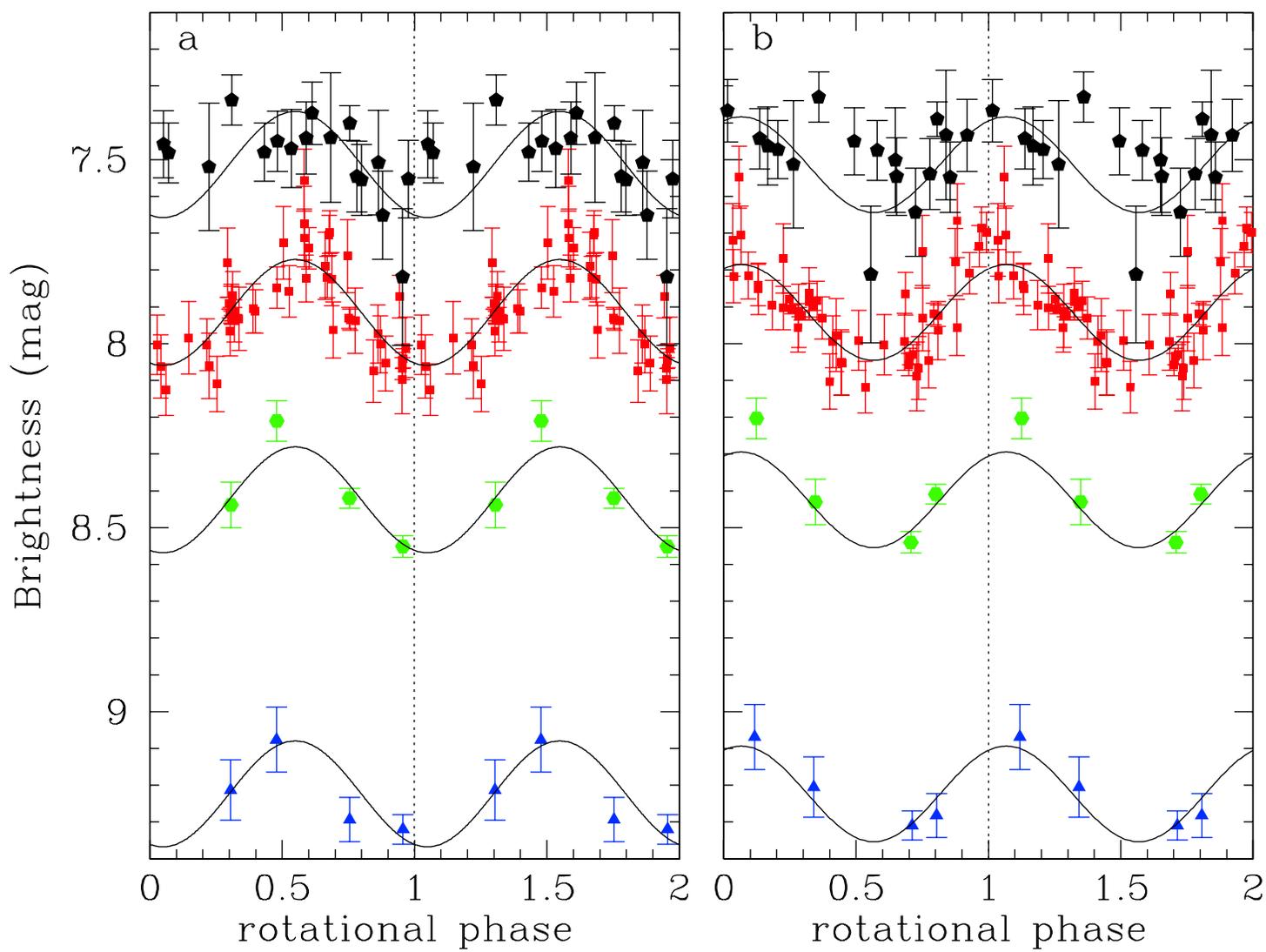

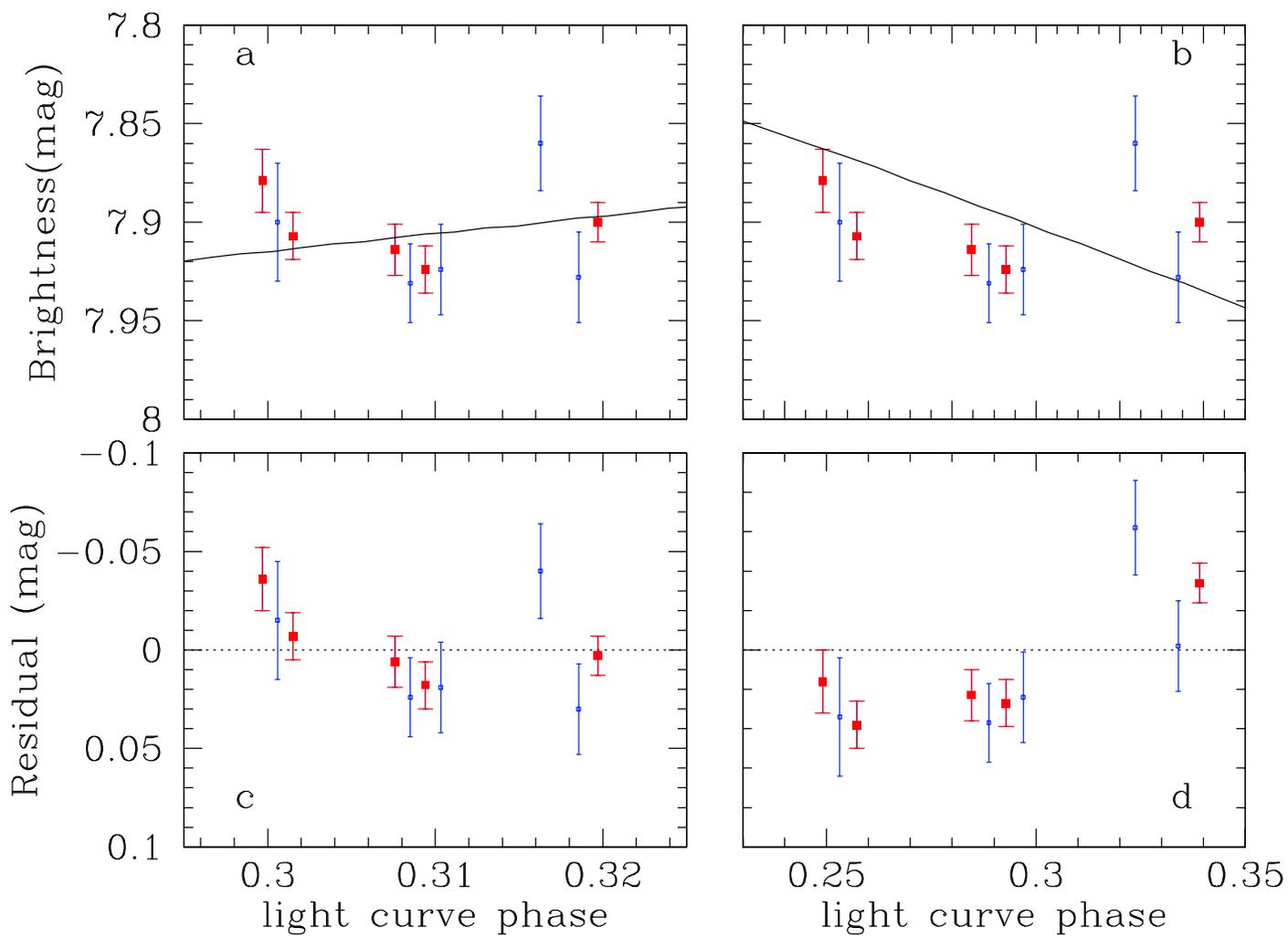

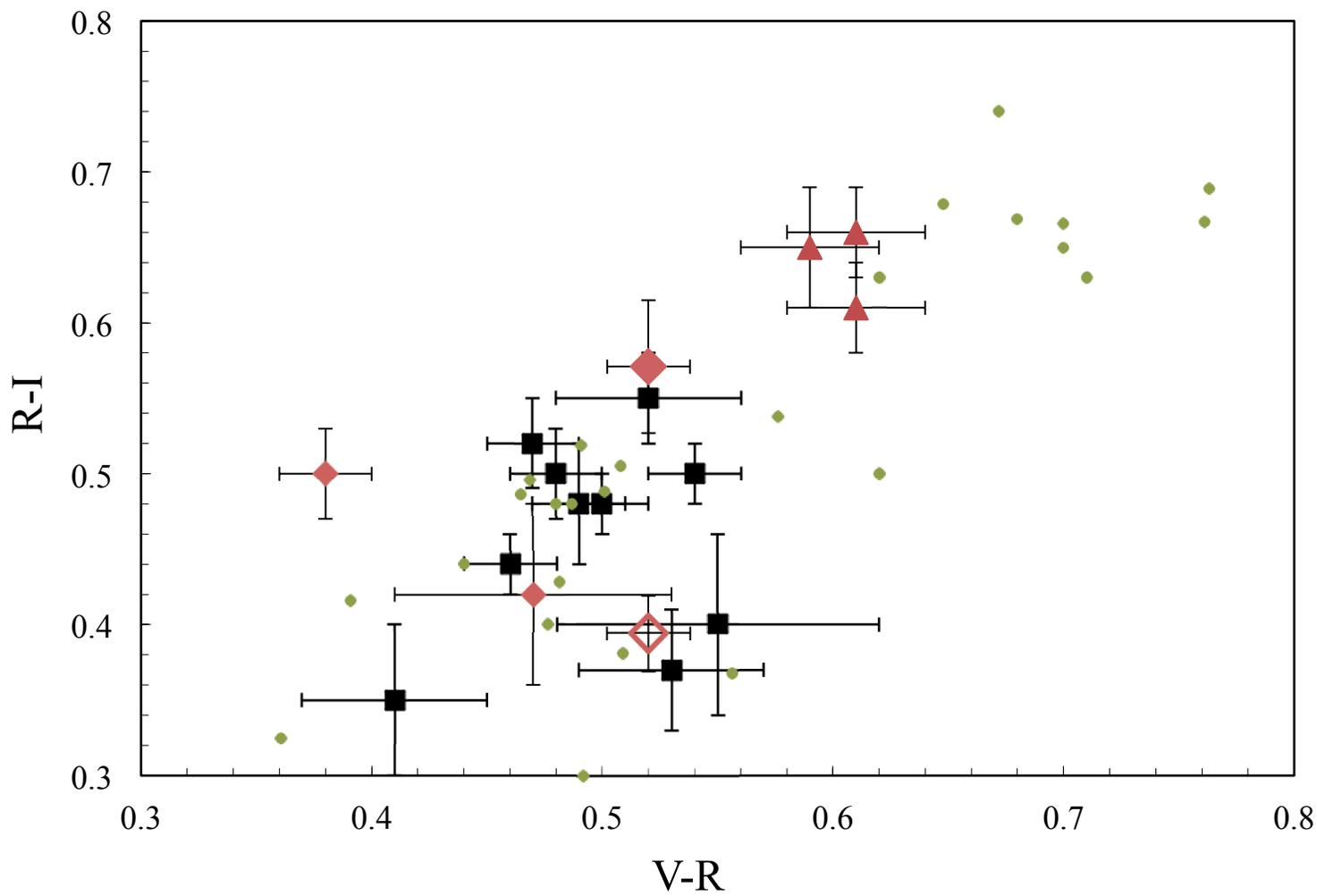